\documentstyle[11pt,epsf,paspconf]{article}

\def\BE{\begin{equation}}
\def\BEL#1{\begin{equation}\label{#1}}
\def\EE{\end{equation}}
\newcommand{\HI}{H\,{\scriptsize I}}
\newcommand{\HII}{H\,{\scriptsize II}}
\newcommand{\Halpha}{H$\alpha$}
\newcommand{\etal}{{\it et al.}}
\newcommand{\eg}{{\it e.g.}}
\newcommand{\ie}{{\it i.e.}}
\newcommand{\cf}{{\it c.f.}}
\newcommand{\COBE}{{\it COBE}}
\newcommand{\IRAS}{{\it IRAS}}
\newcommand{\Rv}{R_V}
\newcommand{\Ab}{A(B)}
\newcommand{\Av}{A(V)}
\newcommand{\Ebv}{E(B-V)}
\newcommand{\kapstar}{\kappa^\star} % Effective cross-section in ISRF
\newcommand{\Tmean}{\langle T \rangle}
\newcommand{\degree}{^\circ}
\newcommand{\cm}{{\rm ~cm}}

\newcommand{\MJypSr}{{\rm ~MJy}/{\rm sr}}
\newcommand{\MHz}{{\rm ~MHz}}
\newcommand{\GHz}{{\rm ~GHz}}
\newcommand{\K}{{\rm ~K}}
\newcommand{\mK}{{\rm ~mK}}

\begin{document}

\title{Interstellar Dust Emission as a CMBR Foreground}
\author{Douglas P. Finkbeiner}
\affil{University of California at Berkeley, Departments of Physics \&
Astronomy, 601 Campbell Hall, Berkeley CA 94720}

\author{David J. Schlegel}
\affil{Princeton University, Department of Astrophysical Sciences,
Peyton Hall, Princeton, NJ 08544}

\begin{abstract}

This chapter discusses Galactic dust and how its thermal emission
confuses CMBR measurements.  Interstellar dust grains are composed
of many differing species, and observational evidence has only begun
to disentagle their composition and sizes.  Fortunately, their
far-IR/millimeter emission is less complex.  We describe how a very
simple two-species model can describe the emission at
$200 < \nu < 3000\GHz$ to high precision.  At lower frequencies,
other non-thermal processes may dominate the emission from dust.

\end{abstract}


\section{Introduction}

\begin{figure}
\epsfxsize=5.0in\epsfbox{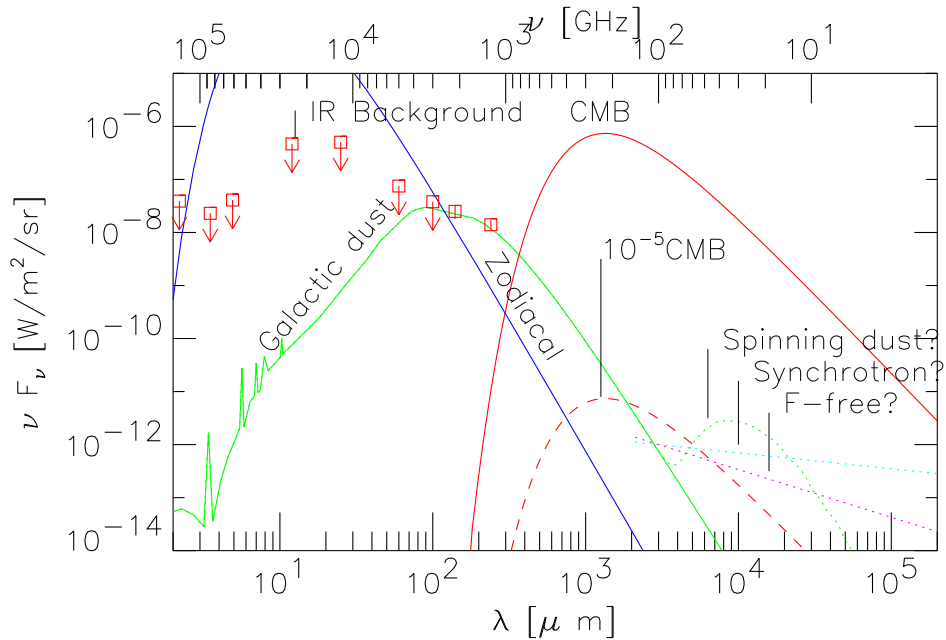}
\caption{Contributions to the infrared/millimeter/radio sky above the
Earth's atmosphere}
\label{fig_nightsky}
\end{figure}

The different sources of radiation in the night sky from the near-infrared
through radio are shown in Figure \ref{fig_nightsky}.  These are to be
compared with the CMBR at a temperature of $2.73\K$, and the fluctuations
in the CMBR which have an amplitude lower by $\sim 10^{-5}$ (dashed line).
On the Wien tail of the CMBR, thermal emission from Galactic dust is
the main source of confusion.  This emission is now well-understood,
thanks to the \COBE/FIRAS experiment (Fixsen \etal\ 1996).
We can predict the thermal emission at all frequencies $\nu \la 3000 \GHz$
to a precision of $\sim 10\%$ (Finkbeiner, Schlegel \& Davis 1999).
This work builds upon the maps of dust column density produced
by Schlegel, Finkbeiner, \& Davis (1998; hereafter SFD98).

At frequencies below $\sim 100\GHz$, there are more possible sources of
confusion with the CMBR.  Galactic dust may be one such source if it is
rotating supra-thermally, as has been proposed by Draine \& Lazarian (1998).
Other possible sources are synchrotron emission,
free-free emission from ionized hydrogen, and radio emission from
extragalactic sources.  We briefly discuss all as possible sources for
anomalous emission seen at $\nu \la 100\GHz$ by CMBR experiments.

%----------------------------------------------------------------------
% FIGURE: Night sky
%----------------------------------------------------------------------

\subsection{Observational Evidence}

% \subsubsection{Optical Manifestations of Dust}

Dark clouds are apparent to any observer who looks at the Milky Way
from a dark site.  These dark regions are especially conspicuous in
the Southern hemisphere where one sees the utter darkness of the Coal
Sack and dark bands across Centaurus and Sagittarius.
William Herschel regarded these regions as curious voids in the
distribution of stars.  In the early 20th century, detailed studies
by Trumpler convincingly demonstrated that interstellar dust was responsible.
Starlight passing through a dusty medium is subjected to both absorption
and scattering.
The combination of these effects is termed extinction, denoted $\Av$
for the magnitudes of visual-band extinction.  Both processes are
wavelength-dependent, resulting in an apparent reddening of the light.
Reddening is usually parameterized by its differential extinction
between $B$- and $V$-band, $\Ebv \equiv \Ab - \Av $.

Advances in infrared astronomy since 1980 have dramatically improved
our understanding of interstellar dust.  Dust scatters some light in
the optical, as can be seen in the Pleiades star cluster.
In our solar system, a thick disk of dust in the zodiacal plane
scatters some sunlight.  However, more energy is absorbed and then re-radiated
in the infrared.  The zodiacal dust re-radiates most of its energy at
$\lambda \ga 10 \micron$.
The energetics of this zodiacal dust would be uninteresting except that
it dominates the sky (above the atmosphere) at wavelengths $\sim 25\micron$.
This zodiacal emission has been extensively modeled by Kelsall \etal\ 1998. 
Interstellar dust is heated to a lower temperature, radiating its energy
at $\lambda \ga 100\micron$.  Approximately $20\%$ of the energy from
starlight in our Galaxy is re-processed this way by interstellar dust.

\subsubsection{Dust Composition}

The diffuse ISM is known to contain many different types of molecules
and dust grains whose optical properties are dissimilar.  Identifications
for some of these constituents have been made from UV absorption features
and mid-IR emission features.  The ubiquitous $2175\AA$ bump (in absorption
from background stars) is most widely attributed to small ($<100\AA$)
graphite particles (\cf, Draine 1989).  Carbon in another form, that of
polycyclic aromatic hydrocarbons (PAHs), is probably responsible for
strong emission features at $3.3$, $6.2$, $7.7$, $8.6$ and $11.3\micron$
(\eg, Allamandola, Tielens, \& Barker 1985; Desert \etal\ 1990).
A review by Henning \& Salama (1998) contains the latest
information on carbon-based grains.
Other mid-IR emission features are attributed to silicates and troilites.
For a coherent and only somewhat dated review, see Whittet (1992). 

\subsubsection{Grain Sizes}

Along with the composition of dust grains, their size distribution
also impacts their extinction and emission properties.
The UV/optical extinction, $A(\lambda)$, is neatly described by a
one-parameter family of curves (Cardelli, Clayton, \& Mathis 1989).
That one parameter is usually chosen as an observable ratio,
\BE \Rv = {\Av \over \Av - \Ab} . \EE
Clayton \& Cardelli (1988) have measured the extinction curve for
76 lines-of-sight in our Galaxy.  Most diffuse clouds favor an $\Rv=3.1$
extinction curve.  However, there are variations in $\Rv$ from $2.6$ to
$5.5$, where the larger values correspond to some dense cloud environments.
Changes in grain size distribution is presumed responsible for these
variations.  On average, interstellar dust grains are thought to follow a
power law distribution of grain sizes $dn/da \propto a^{-3.5}$
(Mathis, Rumpl, \& Nordsieck 1977)
from about $0.0005\micron$ (the size at which absorption of a single photon
can sublimate mass away from the grain) to $0.25\micron$.
The large-size cutoff at $0.25\micron$ is necessary to be consistent
with the measurements of $\Rv$ (Kim, Martin \& Hendry 1994).
Larger grains would be relatively gray absorbers, increasing the value
of $\Rv$ beyond that observed.  So $\Rv$ provides crucial information on
the grain size distribution, and may be found in the future to correlate with
some aspect of the far-IR emission from dust.

\subsubsection{Polarization Effects}

Besides attenuating starlight, dust grains are known to polarize that
light.  To date, polarization of optical light has been measured
towards $\sim 10,000$ stars (Mathewson \& Ford 1970; Korhonen \& Reiz
1986), some high above the Galactic plane (Berdyugin \& Teerikorpi
1997).  Regions of higher polarization are correlated with regions of
higher extinction (\eg, Larson, Whittet, \& Hough 1996).  One expects
polarization only if the dust grains are non-spherical and have a
preferential axis of alignment.  For this reason, polarization studies
have been used to argue for coherent, large-scale magnetic fields in
the Galaxy (\eg, Heiles 1996).  The magnetic moments of the dust grains
will preferentially align with the ambient magnetic field.
As these aligned grains rotate, they emit polarized radiation
that may interfere with attempts to measure the polarization of the CMBR
anisotropy. 

\subsubsection{Dust Distribution}

The morphology of the interstellar dust is quite complex, consisting
of dark clouds and long filamentary structures arching across the sky.
The diffuse, wispy parts of this emission are known as cirrus, to
differentiate them from more compact features (often associated with
individual objects).  Not surprisingly, the dust
emission is concentrated in the Galactic plane, but significant cirrus
is observed in every part of the sky at $100\micron$ 
(Low \etal\ 1984).  SFD98 have computed the power spectrum of this dust
at latitudes $|b| > 45\degree$, finding a reasonable fit to
$P(k)\propto k^{-2.5}$.  However, as this cirrus is in coherent structures, a
1-D power spectrum is not statistically sufficient to describe its distribution.
This morphology is another obstacle to CMBR measurements since it is not at
all reminiscent of gaussian random-fields that CMBR pundits usually model.
However, this phase coherence provides avenues that
may prove helpful in determining the quality of a dust foreground
model as applied to real microwave data.  For example, we can easily
measure the 3-point function of the Galactic dust.  If the CMBR is
presumed to be gaussian, a non-zero 3-point function in a
foreground-subtracted CMBR map will measure the level of residual
contamination (Spergel 1998).

\subsection{Far-IR Emission}

In spite of the expected melange of dust grains, it was originally
expected that in the far-IR/submillimeter bands, all dust would have
similar optical properties.  For example, Draine \& Lee (1984)
predicted $\nu^2$ emissivity for both silicate and graphite grains
whereby the emission is described by
\BE I_\nu = B_\nu(T) \nu^2 . \EE
The theoretical motivation for this emissivity is based on the fact that the
response of a damped harmonic oscillator to a driving force varies as
$\nu^2$ in the low frequency limit (far below resonance).  Even though
microwave frequencies are below all of the phonon modes in the
smallest crystals, this is not true for grains bigger than $\sim
0.1\micron$, and these bigger grains are expected to dominate the
emission.  More recent laboratory measurements (see Pollack \etal\
1994 for a review) suggest that this is an oversimplication, with
different species of grains having differing emissivity laws in the
far-IR.  In particular, glassy materials (e.g. amorphous silicates) in
some cases radiate more efficiently than their crystalline
counterparts, and are more closely characterized by a $\nu^{1.5}$
emissivity (Agladze \etal\ 1996).  As we show below, these silicates
probably dominate the millimeter/microwave emission relevant to CMBR
studies.

Fortunately, complete knowledge of the size and composition of interstellar
dust grains is not necessary to predict their impact upon CMBR studies.
% The far-IR/submillimeter dust emission is dominated by large grains.
Since the long wavelength emissivity of a grain scales as its size $a$
times the surface area, or volume $a^3$, the larger grains dominate
the submillimeter emission (Draine \& Anderson 1985)
The very small grains (VSGs) are transiently heated to higher temperatures
and dominate the dust emission at $\lambda \la 100\micron$, but do not
contribute significantly to submillimeter emission.
The opacities of dust grains are exceedingly complicated in the UV/optical
where they absorb energy, and slight changes in composition can significantly
alter them.  However, their far-IR opacities are not nearly as sensitive.
Both laboratory and astronomical measurements are well-fit at
$\lambda \ga 1000 \micron$ ($\nu \la 300 \GHz$) using power-law
emissivities with indices in the range from $1$ to $3$
(see Pollack \etal\ 1994).

\section{Microwave Foregrounds}

Interstellar dust is only one of the many foregrounds that complicate
measurement of the CMBR anisotropy.  An analysis of the relative
importance of the various foregrounds is given by Tegmark \&
Efstathiou (1996).  In this section, we provide a brief overveiw of
the most important foregrounds. 

At low frequencies, the dominant diffuse Galactic emission is due to
synchrotron radiation produced by relativistic electrons interacting
with the Galactic magnetic field.  A map at 408 MHz by Haslam \etal\
(1981) is the only full-sky synchrotron template in existence.
Reich \& Reich (1986) cover the northern sky in the $1420 \MHz$
continuum, and are complemented by the new Rhodes/HartRAO survey at
$2326 \MHz$ in the south (Jonas, Baart, \& Nicolson 1998).  Where these
maps all overlap, they allow examination of the synchrotron radiation
at 3 frequencies -- and indicate significant spectral index variation
across the sky.  The synchrotron radiation is usually characterized by
a spectral index $\beta=-2.7$ in units of brightness temperature, or
$\beta=-0.7$ in units of flux (\eg, MJy/sr).  The typical
high-latitude brightness temperature at $20 \GHz$ predicted from these
data is approximately $0.5 - 1 \mK$, rising to no more than $20 \mK$ in the
Galactic plane.  Extrapolating the power law to $100 \GHz$, the
synchrotron emission would contribute at most 10 $\mu K$ to the
emission at high latitude.  However, the index $\beta$ is seen to
steepen to $-2.9$ or less even at low frequencies, due to a cutoff in
the electron energy distribution (Jonas 1998). Therefore,
synchrotron emission is unlikely to be an important emission component
at high latitude at $\nu \ga 30 \GHz$.

Another diffuse component that is not well constrained is free-free
emission from ionized gas.  Except in \HII\ regions, free-free is lost
in the signal of synchrotron emission at low frequencies (\eg, at 408 MHz).
Although there is known to be ionized gas at high latitude from
\Halpha\ emission (Haffner, Reynolds, \& Tufte 1998), so far no
convincing correlation has been seen between \Halpha\ data and $\sim
20 \GHz$ data.  As more data of both types become available, we expect
to see a correlation.  Another possible emitter of free-free is the
Local Hot Bubble (LHB) of ionized gas in which the Sun and many other
stars reside.  A map of this bubble was produced using the absorption
of soft X-rays observed by ROSAT (Snowden \etal\ 1998).  Because the gas in
the LHB is very hot, the anticipated \Halpha\ emission per microwave
emission is very low, so this will not appear in the \Halpha\
template.

In addition to these diffuse components, a wealth of point sources are
apparent at radio and far-IR frequencies.  Several hundred thousand
radio sources have been observed in the PMN (Griffith \& Wright 1993)
and NVSS (Condon \etal\ 1998) surveys.  At $60\micron$, SFD98
identified $\sim 20,000$ extragalactic sources and unresolved Galactic
sources.  These sources will contribute a great deal of non-gaussian
contamination to the CMBR.  The brightest ones are very easy to
identify and remove.  Preliminary work has shown that removing the
brightest sources will be sufficient for power-spectrum analyses with
the MAP experiment (Tegmark \& de Oliveira-Costa 1998).  Point source
decrements due to the Sunyaev-Zeldovich effect will have also been
seen, and their impact on CMBR observations has been noted (Aghanim
\etal\ 1997). 

As we have emphasized,
the dust emission described in the next section is not the only
diffuse foreground, but it is the brightest at high frequencies
($\nu \ga 100\GHz$).  However, it is important for another reason.  The
excess dust-correlated emission observed in the DMR $31.5$ and $53 \GHz$
channels (Kogut \etal\ 1996) and later by other experiments
(e.g. Leitch \etal\ 1997; de Oliveira-Costa \etal\ 1998) is still
unexplained.  It may be caused by dust-correlated free-free emission.
On the other hand, it may be the result of rapidly rotating dust
grains (Draine \& Lazarian 1998).  Whatever this component is, it appears to 
correlate fairly well with dust, making the determination of a good
dust template more important than thermal (vibrational) dust emission
alone would indicate. 

\section{Single Component Dust Model}

A simple but naive prediction for millimeter/microwave dust emission
can be made from our previous work.  SFD98 extensively studied the
emission from dust in the regime $100\micron < \lambda < 240\micron$.
Assuming a $\nu^2$ emissivity model, the temperature of the dust was
mapped with a resolution of $1.3\degree$ from the DIRBE
$100\micron$/$240\micron$ emission ratio.  The $100\micron$ emission
of the dust was mapped with a resolution of $6.1\arcmin$
by utilizing small-scale information from the \IRAS\ Sky Survey Atlas
(Wheelock \etal\ 1994).
Emission at lower (millimeter/microwave) frequencies can be predicted by
extrapolating the $100\micron$ flux using this temperature fit.
For each line-of-sight in the maps, the emission at frequency $\nu$ can
be expressed as
\BEL{equ_extrap1}
    I_\nu = K^{-1}_{100}(\alpha,T) I_{100}
    {\nu^\alpha B_\nu(T) \over \nu_0^\alpha B_{\nu_0}(T) }
\EE
where $B_\nu(T)$ is the Planck function at temperature $T$,
$I_{100}$ is the DIRBE-calibrated $100\micron$ map, $K_{100}(\alpha,T)$
is the color correction factor for the DIRBE $100\micron$ filter when
observing a $\nu^\alpha$ emissivity profile, and $\nu_0=3000\GHz$ is
the reference frequency corresponding to $100\micron$.

\begin{figure}
\epsfxsize=5.5in\epsfbox{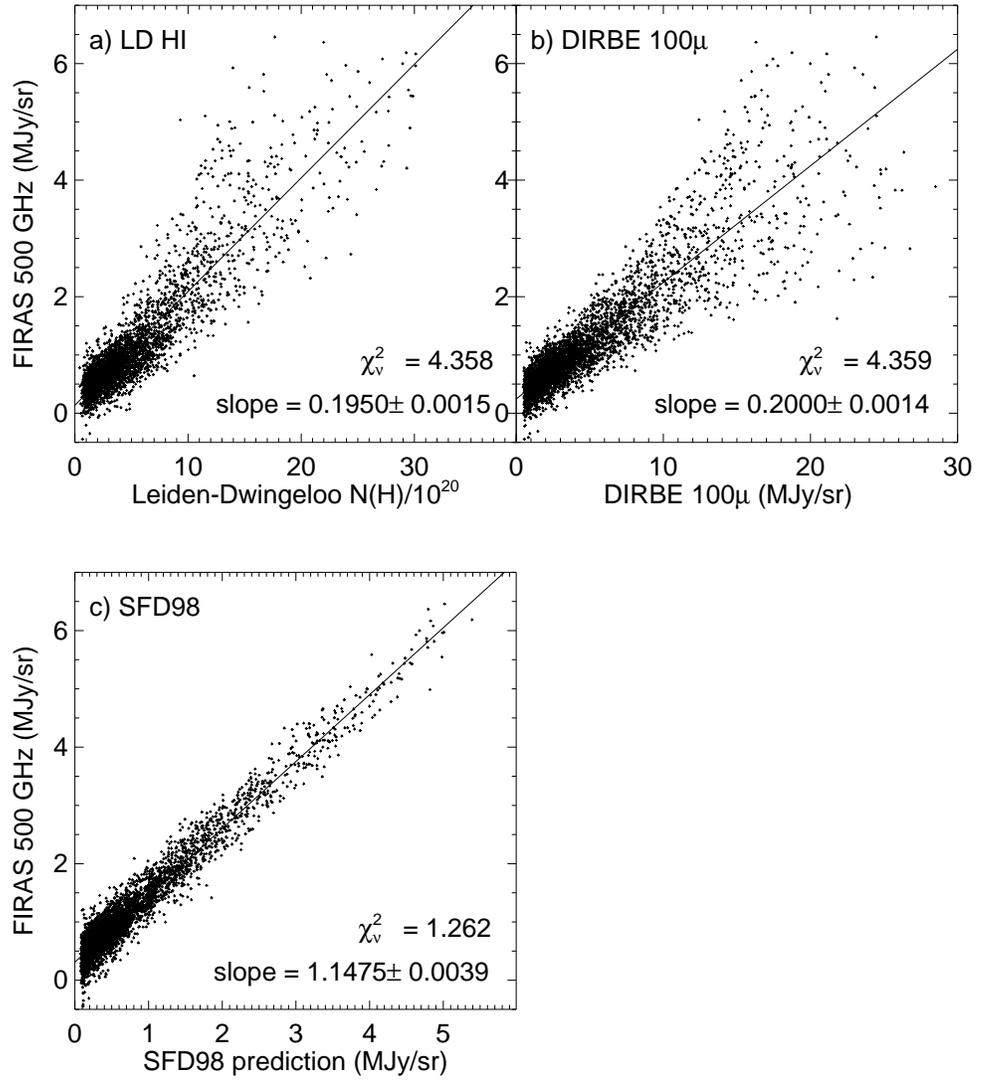}
\caption{Comparison of the FIRAS data at $500\GHz$ with three predictors.}
\label{fig_firas_scatter}
\end{figure}

The extrapolated, millimeter emission from dust as predicted from
SFD98 can be compared directly to the FIRAS measurements.
We synthesize a broadband FIRAS $500 \GHz$ channel (to increase S/N)
and show that the correlation with predictions from equation \ref{equ_extrap1}
is very good (Figure \ref{fig_firas_scatter}c).
For comparison, we plot the correlation
with (a) \HI\ column density, and (b) DIRBE $100\micron$ flux.
Both of these show a scatter that is 3.7 times worse than with SFD,
demonstrating that millimeter emission from dust is neither simply related
to the \HI\ column density, nor is the dust at one temperature everywhere
on the sky.

%----------------------------------------------------------------------
% FIGURE: FIRAS scatter
%----------------------------------------------------------------------

\begin{figure}
\epsfxsize=4.5in\epsfbox{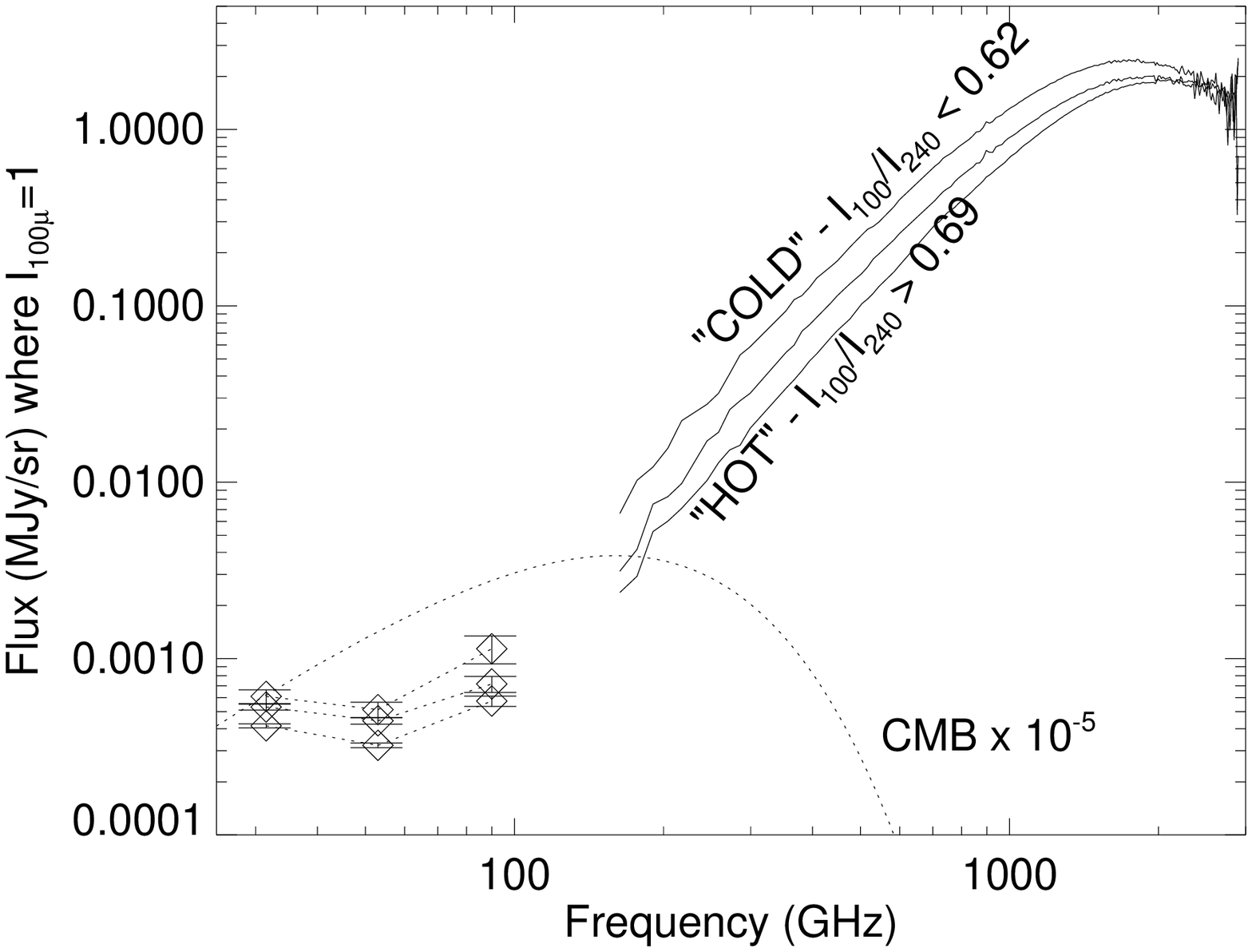}
\caption{Difference spectra from the DMR ($\nu < 100\GHz$) and
DMR ($\nu > 200\GHz$) instruments.  These spectra divide the sky into
three regions based upon temperature, and are renormalized to
$I_{100\micron} = 1\MJypSr$. }
\label{fig_3temp}
\end{figure}

Our regression line in Figure \ref{fig_firas_scatter} differs significantly
from unity, indicating that a $\nu^2$ emissivity is incorrect for the dust.
At lower frequencies, the slope departs even more strongly from unity.
This is an indication that a $\nu^2$ emissivity is incorrect for the dust,
as can be seen from the mean spectrum of large regions of the sky
(Figure \ref{fig_3temp}: since the DMR is a differential instrument,
we have plotted the difference between bright and faint regions of the sky.)

%----------------------------------------------------------------------
% FIGURE
%----------------------------------------------------------------------

The SFD prediction can be made using other emissivity profiles by
modifying the exponent, $\alpha$, in equation \ref{equ_extrap1}.
An $\alpha=1.5$ emissivity profile results in a better fit
at low frequencies, but ruins the fit at high frequencies.
In fact, there is \emph{no} power-law emissivity model that adequately
fits the FIRAS data.

\section{Multi-Component Dust Model}

Motivated by observations of many different grain species in the ISM,
we have formulated a general description of a multi-component ISM
suitable for microwave predictions (Finkbeiner \etal\ 1999).  We
capitalize upon the uncertain degeneracies between components to
construct a multi-component model with very few degrees of freedom.
As we discussed, any one component of dust might be expected to
exhibit power-law emissivity in the range $100 \la \nu \la 1000 \GHz$.
The complicated behavior of their optical opacities is unimportant.
The only relevant optical parameter is $\kapstar$, the effective
cross-section to the interstellar radiation field (ISRF): \BE
\kapstar_k = { \int_0^{\infty} \kappa_k(\nu) E(\nu) d\nu \over
\int_0^{\infty} E(\nu) d\nu } \EE where $E(\nu)$ represents the
angle-averaged intensity in the ISRF.

We construct a multi-component model of the ISM, analogous
to equation \ref{equ_extrap1}, by summing the emission of these components:
\BEL{equ_multi}
   I_{\nu} =
   { \sum_k{ f_k \epsilon_k(\nu) B_\nu(T_{k})}
    \over \sum_k{ f_k \epsilon_k(\nu_0) B_{\nu_0}(T_{k})
        K_{100}(\alpha_k,T_{k})} } I_{100}
\EE
where $f_k$ is the mass fraction of the $k$-th component, $T_{k}$ is
the temperature, $K_{100}$ is the DIRBE
color-correction factor and $I_{p,100}$ is the SFD98
$100\micron$ flux.
The mass fractions are forced to sum to unity, \ie\ $ \sum_k f_k = 1 $.
Each emissivity is a power-law with index $\alpha_k$.
The observationally relevant parameter is the ratio of far-IR emissivity
to the effective optical opacity, $q = \epsilon(\nu_0)/\kapstar$.

Each dust component is therefore described by three global parameters
($f_k$, $q_k$, $\alpha_k$) and one parameter
that varies with position on the sky, \eg, $T_1(\vec{x}$).
The temperatures of all dust species are coupled by demanding that they
all respond to the same ISRF (for more details, see Finkbeiner \etal\ 1999).
Whether or not the actual components of the dust physically correspond to these
components, equation \ref{equ_multi} can be thought of as a phenomenological
``expansion set'' for describing the composite dust spectrum.

Remarkably, we find that only two components of dust are required to
describe the spectrum everywhere on the sky as seen by FIRAS.
As a first guess, we tested a two-component model designed to replicate
the spectrum in Pollack \etal\ 1994.
The prescription in their paper for their best-fit broken
power law corresponds in our model to $\alpha_1=1.5, \alpha_2 = 2.6,
f_1=0.25, $ and $q_1/q_2=0.61$.  This results in a considerably better
fit without fitting any new free parameters.
% of $\chi^2_\nu= 16.6$
% The ratio of the Pollack \etal\ model to the strawman $\nu^2$ model is
% overplotted in Figure \ref{fig_firas_slope_spec} as a solid line.
Between $800$ and $1800 \GHz$, where the FIRAS signal is very good, the
model matches the data to approximately 1\% everywhere.
At lower frequencies, the largest deviation is $10\%$.

Allowing $f_1$ and $q_1/q_2$ to float provides even better fits.
Our best fit is achieved with two components with power-law emissivities
$\alpha_1 = 1.68$, $\alpha_2 = 2.78$, where the former component
dominates the emission at $\nu \la 500 \GHz$. 
We have fit 71\% of the sky as observed by FIRAS (4378 pixels)
at 123 frequencies, for a total of $\sim 540,000$ data value comparisons.
We also effectively fit for an unconstrained zero-point value at each
frequency to avoid having the solution influenced by uncertainties in
the cosmic infrared background.
The resulting reduced $\chi_\nu^2$ is $2.3$/DOF, as compared to $\sim 30$ (!)
for the best single-component model.  Results for a suite of models,
including our best fit 2-component model, are given in Finkbeiner \etal\ (1999)
along with tests of their robustness. 

%----------------------------------------------------------------------
% FIGURE: Leitch et al data
\begin{figure}
\epsfysize=7.0in\epsfbox{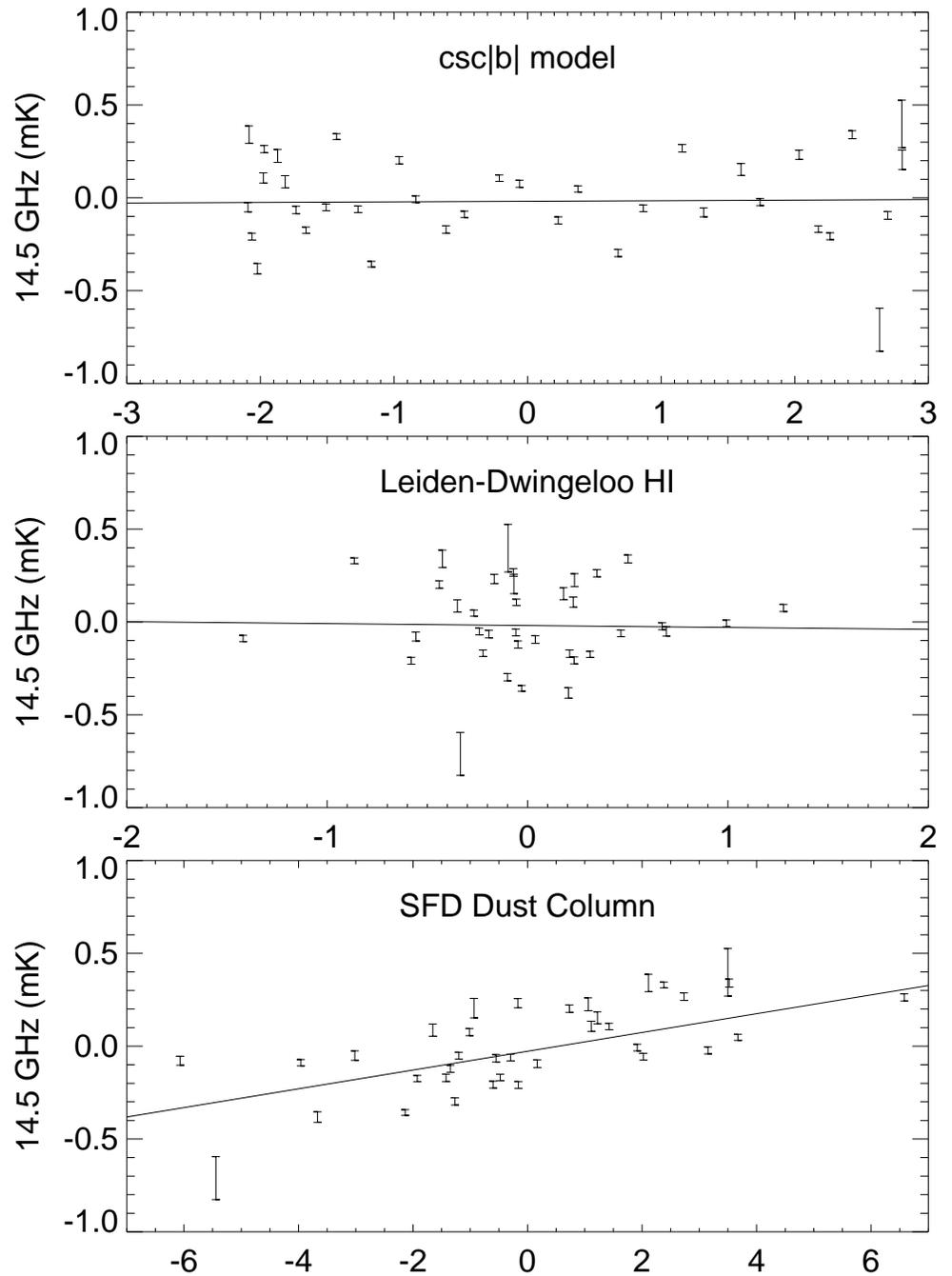}
\caption{Correlation of different templates with the Leitch \etal\ 1997
CMBR data.  A $\csc|b|$ model or 21-cm map (from Leiden-Dwingeloo)
are poor predictors for microwave dust emission.}
\label{fig_leitch}
\end{figure}
%----------------------------------------------------------------------

\section{Discussion}

The thermal emission from Galactic dust can be very successfully
predicted at millimeter/microwave frequencies using a two-component
composition model with temperature varying on the sky.  We interpret
the best fit as a amorphous silicate-like component ($\nu^{1.7}$
emissivity, $\Tmean \approx 9.5\K$) and a carbonate-like component
($\nu^{2.8}$ emissivity, $\Tmean \approx 16\K$).  The FIRAS data are
also reasonably well fit using two $\nu^2$ emissivity components, if one of
these components is very cold ($\Tmean \approx 5\K$) as proposed by
Reach \etal\ (1995).  We find no evidence for a warm component ($\nu^1$
emissivity, $\Tmean \approx 29\K$) as proposed by Lagache \etal\ 1999.

A full range of Galactic dust models are discussed in Finkbeiner
\etal\ 1999.  Numerical predictions for thermal emission will be made
available at our web page,
\begin{quote} {\tt http://astro.berkeley.edu/dust} , \end{quote}
or at the FORECAST (FOREgrounds and CMB Anisotropy Scan simulation
Tools) home page,
\begin{quote} {\tt
http://cfpa.berkeley.edu/forecast}.
\end{quote}
By using the \IRAS\ maps as reprocessed by SFD98, these predictions
will be provided at a resolution of $6{\farcm1}$.

It is critical that CMBR experimentalists compare their observations
with valid models for the Galactic dust emission.  A ``template approach''
is often carelessly 
used to compare observations with alleged contaminants, with
the correlation amplitude indicating the level of contamination.
For example, radio maps (\eg, the Haslam \etal\ map at $408 \MHz$)
are used as templates for synchrotron emission at much higher
frequencies, in spite of the well known spectral index variation. 
$21\cm$ maps (\eg, Leiden-Dwingeloo)
are often used as templates for microwave dust emission.
This can be a terrible template, as it ignores variations in the
dust/gas ratio and the well-measured variations in dust temperature.
This point is demonstrated in Figure \ref{fig_leitch}, where we
have correlated the Leitch \etal\ (1997) 14.5 GHz data with possible templates
for the dust emission.  The microwave data shows very little correlation
with a $21\cm$ template, or with the even more simple-minded (but still
occasionally used) $\csc|b|$ model for Galactic dust.  
One might incorrectly conclude from this
that there is little dust contamination in this CMBR data.
In fact, when we correlate the Leitch \etal\ data with a proper prediction
at $14.5 \GHz$, there is evidence for significant contamination.  A
naive comparison with $100\micron$ IRAS flux also indicates a strong
correlation with dust - but with the wrong amplitude.  We find that
the dust on the $\delta=88\degree$ ring used by Leitch \etal\ is cold
enough to lower the $100\micron$ emission per dust column by a factor
of 2.75 relative to the average high-latitude ratio.  Therefore, the
amount of 14.5 GHz emission per dust column is overstated by a factor
of $\sim3$ in comparisons such as the famous Figure 11 in Draine \&
Lazarian (1998).  We hope that our dust spectrum model will allow both a
realistic prediction of the thermal (vibrational) dust emission, and
also facilitate a meaningful comparison between the dust and the
mysterious excess 15-90 GHz emission component that correlates so well
with it. 

%------------------------------------------------------------------------------

\acknowledgments

The \COBE\ datasets were developed by the NASA Goddard Space Flight
Center under the guidance of the \COBE\ Science Working Group and were
provided by the NSSDC.  Dale Fixsen has been very helpful in understanding
the calibration uncertainties in the \COBE/FIRAS data.

%------------------------------------------------------------------------------

\end{document}